\documentclass{rmaa}
\usepackage{astro_bib_macro}
\usepackage{natbib}
\bibpunct{(}{)}{;}{a}{}{,}

\title{Modelling of asymmetric nebulae. II. Line profiles.}

\author{C.~Morisset \altaffilmark{1}
   and G.~Stasi\'nska \altaffilmark{2}
}

\altaffiltext{1}{Instituto de Astronom\'{\i}a, UNAM, Morisset@Astroscu.UNAM.mx}
\altaffiltext{2}{LUTH, Observatoire de Meudon}

\shortauthor{Morisset \& Stasinska}
\shorttitle{Modelling of nebular line profiles}

\fulladdresses{
\item C. Morisset: Instituto de Astronom\'{\i}a,
   Universidad Nacional Aut\'onoma de M\'exico
   Apdo. postal 70--264; Ciudad Universitaria;
   M\'exico D.F. 04510; M\'exico.
\item G.~Stasi\'nska : LUTH, Observatoire de Meudon, 5 place J. Janssen,
   F-92195 Meudon Cedex, France.
}

\newcommand{\comment}[1]{}
\newcommand{\news}[1]{#1}
\newcommand{\mysp}{}\def\mysp/{}

\newcommand{\alloa}[3]{\ion{#1\mysp/}{#2}\ #3\AA}
\newcommand{\forba}[3]{[\ion{#1\mysp/}{#2}]\ #3\AA}

\newcommand{\uchii}{}\def\uchii/{UCHII}
\newcommand{\teff}{}\def\teff/{$\mathrm{T}_{\mathrm{eff}}$}
\newcommand{\kms}{}\def\kms/{km s$^{-1}$}

\newcommand{\nebu}{}\def\nebu/{NEBU\_3D}
\newcommand{\hbeta}{}\def\hbeta/{H$\beta$}

\abstract{

We present a tool, VELNEB\_3D, which can be 
applied to the results of 3D 
photoionization codes to generate emission line 
profiles, position-velocity maps and 3D maps in 
any emission line by assuming an arbitrary 
velocity field.

We give a few examples, based on our pseudo-3D 
photoionization code NEBU\_3D \citep{MSP05a} 
which show the potentiality and usefulness of our 
tool. One example shows how complex line profiles 
can be obtained even with a simple expansion law 
if the nebula is bipolar and the slit slightly 
off-center.
Another example shows different ways to produce 
line profiles that could be attributed to a 
turbulent velocity field while there is no 
turbulence in the model. A third example shows 
how, in certain circumstances, it is possible to 
discriminate between two very different 
geometrical structures -- here a face-on blister 
and its ``spherical impostor'' -- when using 
appropriate high resolution spectra. Finally, we 
show how our tool is able to generate 3D maps, 
similar to the ones that can be obtained by 
observing extended nebulae with integral field 
units.

}

\addkeyword{methods: numerical}
\addkeyword{line: profiles}
\addkeyword{turbulence}
\addkeyword{ISM: H II regions}
\addkeyword{planetary nebulae}

\begin{document}

\maketitle

             \section{Introduction}
\label{sec:intro}

The study of emission line profiles in nebulae 
(planetary nebulae and HII regions) is important 
for many reasons. The shapes of emission lines 
result from the motions along the line of sight 
of the parcels of gas covered by the 
observational aperture and in which these lines 
are emitted. With the help of some hypothesis, 
the analysis of line profiles allows one to 
describe the internal motions in nebulae. It also 
opens access to the third dimension in the study 
of nebulae, since an analysis of surface brightness 
profiles alone may be misleading \citep[see 
e.g.][]{MSP05a}. A proper description of the 
geometry and kinematics of nebulae is a 
prerequisite to understand their dynamical 
evolution since the epoch of their formation. 
Unfortunately, the interpretation of line 
profiles is extremely difficult. The first 
spectra that showed line splitting in planetary 
nebulae were published by 
\citet{1918PLicO..13...75C} and were interpreted 
as due to rotation. The interpretation in terms 
of nebular expansion, which physically makes more 
sense, appeared only in 1931 
(\citeauthor{1931ZA......2..329Z}). An extent 
discussion of internal motions in planetary 
nebulae appears with the work of 
\citet{1950ApJ...111..279W}. Among other things, 
he noted that the proportion of nebulae with 
splitted lines in which the blue component is 
stronger than the red component is equal to that 
of nebulae in which the reverse is occurring, 
which implies that the nebulae are transparent in 
the observed spectral range. He concluded that 
the relative strengths of the red and blue 
components depend upon non uniform distribution 
of the nebular material and upon the direction of 
the observer. He also noted that the separation 
between the two components is smaller for 
[\ion{Ne}{v}] than for [\ion{O}{iii}] and 
[\ion{Ne}{iii}] and concluded that the expansion 
velocity of the nebular material increases with 
distance to the star. This however was shown 
later not to be a general property of planetary 
nebulae \citep{1982AA...109..131S}. 
\citet{1966ApJ...145..697O} observed emission 
line profiles in several bright planetary 
nebulae, and noted that the lines are broader 
than the expected thermal Doppler widths, 
indicating significant mass motions within the 
nebulae. \citet{1958RvMP...30.1025W} and then 
\citet{1968ApJ...153...49W} were, to our 
knowledge, the first to perform a 
spatio-kinematic analysis of the  data under the 
assumption of ellipsoidal geometry. For the five 
nebulae he studied, Weedman found that the 
expansion velocity increases with distance from 
the star, and that the gradient extrapolates to 
zero velocity at a point between the nebular 
shell and the center. This implies that the 
expansion velocity has increased since the time 
of ejection.

Since then, spatio-kinematic studies of planetary 
nebulae have gone into several directions. One is 
a purely empirical approach which, from 
observations of splitting, width and intensity 
ratio maps of emission lines infers the nebular 
structure, using a few simplifying assumptions 
\citep{1985MNRAS.217..539S}: this technique has 
been called tomography. Other studies have 
focused on the kinematics of microstructures 
\citep[e.g.][]{1993ApJ...411..778B,1998MNRAS.294..201M}. 
On the other hand, \citet[][and references 
therein]{GAZ03,GZ03} have determined the internal 
velocity fields of about 70 planetary nebulae, by 
comparing observed profiles with profiles 
computed for  1D photoionization models, while 
\citet{MGV00} have developed a tool to derive 
morphological and kinematical information on 
planetary nebulae, based on a 3D photoionization 
code.

But 3D photoionization codes are heavy in terms 
of computing time, while interpretations of 
kinematic observational data for real planetary 
nebulae with a 1D tool can be misleading.
Here, we present a tool, VELNEB\_3D, and its 
association to a quick pseudo-3D photoionization 
code NEBU\_3D \citep{MSP05a} which allows one to 
easily obtain realistic emission line profiles 
and radial velocity maps for any geometry.

The main goal of the present paper is to emphasize a few points:
\begin{itemize}
\item Emission line profiles are a combination of a morphology of the emitting gas and a velocity field. The complexity of  observed line profiles can be attributed to a complex velocity law (as has been done by Gesicki, Zijlstra et al. for years, using simple non-realistic spherical geometry) or to a complexe morphology (as commonly observed).

\item Velocity fields can be partially attributed to turbulence (as claimed by Gesicki \& Zijlstra, 2003, for a number of planetary nebulae) simply because the analysis unduely relies on spherical models. If the models account for departure from sphericity (as ours do), there is noo need of turbulence to account for the observed profiles

\item Line profiles can allow to distinguish between very different morphologies (a blister and its spherical impostor)
\end{itemize}

The structure of the paper is the following. 
Section ~\ref{sec:velneb3d} presents the 
VELNEB\_3D tool \news{and compares  it to 
previous similar tools.} The three next sections 
present a few applications. The main goal of the 
present paper is to present the capacities of 
this tool and no attempt is made to reproduce 
observations of any real object. 
Section~\ref{sec:appli1} illustrates how complex 
emission line profiles can be obtained using 
simple expansion laws. 
Section~\ref{sec:turbulence} demonstrates how 
aspherical geometries can, in certain 
circumstances, lead to a broadening of emission 
lines that can be mistaken for turbulence. 
Section~\ref{sec:blister} shows an example when 
two very different geometries can be 
distinguished using emission line profiles. 
Section 6 shows an example of a 3D map in several 
lines generated by our code. Section 7 presents a 
summary and some prospects.

             \section{VELNEB\_3D}
\label{sec:velneb3d}
            \subsection{Description}
VELNEB\_3D (which has been developed in IDL) uses 
the results of a 3D photoionization code. In this 
paper, we use the pseudo-3D photoionization code, 
\nebu/. Both
\nebu/ and the visualisation tool VISNEB\_3D have 
been extensively described by \citet{MSP05a}. In 
this section we briefly recall how they operate. 
\nebu/ is a set of tools that allows the user to 
construct a 3D photoionized model nebula from a 
set of runs of a 1D photoionization code 
\citep[in this paper, the 1D code used is NEBU, 
see][]{mp96,P02}. The interpolation is done on a 
coordinate cube for all the relevant physical 
parameters (electron temperature, density and 
ionic fractions) and for the emissivities of 
emission lines the user could need. After the 
construction is done, VISNEB\_3D is used to 
rotate the nebula to any orientation, and to 
compute emission line intensity maps integrated 
along any line of sight.

One can then attribute a velocity vector to each 
cell of the coordinate cube containing the 
nebula. For example, one can adopt a simple 
radial expansion velocity law, where the 
magnitude of the velocity is proportional to the 
distance to the star, like in a Hubble flow. In 
each cell of the cube, the code computes 
elementary emission line profiles, using the 
velocity vector defined previously and taking 
into account the thermal broadening obtained from 
the local electron temperature and the mass of 
the emitting ion. A turbulence velocity can be 
quadratically added to the thermal velocity. An 
integration of the individual emission line 
profiles along a line of sight is performed for 
each pixel of the surface brightness map. The 
procedures used here are very similar to the ones 
described in more detail in \citet{MGV00}.

Aperture effects can be simulated by applying a 
mask corresponding to a slit size, orientation 
and shift relative to the center of the image. 
The effect of the seeing is taken into account by 
smoothing the aperture mask by a given angular 
size square kernel.
The resulting profile can also be finally 
convolved by a theoretical instrumental profile.

All this constitutes the VELNEB\_3D tool, which 
is here applied to results from a photoionization 
calculation with NEBU\_3D, but can be applied to 
results from any photoionization code. As a 
matter of fact, VELNEB\_3D in itself is likely 
very similar to the spatio-kinematic code 
mentioned by \citet{2004MNRAS.348.1047H}. The 
fact that VELNEB\_3D is coupled to a 
photoionization code ensures that the line 
profiles are computed in a consistent way, taking 
into account the distribution of ions and 
temperature throughout the nebula rather than 
arbitrarily assuming a source function for each 
line.

             \subsection{Comparison with similar tools}
\news{We now compare the relative merits and 
drawbacks of several recent tools that have been 
designed to study nebular expansion.

\citet{1993ApJ...404L..25F} compute 2D fully hydro simulations including photoionization. Synthetic position-velocity maps are computed and compared by eye with real observations, showing that the 2-wind model accounts qualitatively for the diversity of PN shapes and velocity fields. This is probably the most elaborate approach to study the morpho-dynamics of planetary nebulae. 
\citet{2005AA...431..963S} and \citet{2005AA...441..573S}
develop a 1D hydromodelling code, including photoionization. They compare surface brightness distributions and line profiles to observations of a few planetary nebulae. One of the big limitation of these works is the spherical morphology. 
\citet{2005AA...444..849R} performe three-dimensional AMR simulations of point-symmetric nebulae and compute PV-diagrams, but only for the hydrogen recombination lines.
In addition, both this code and the \citet{2005AA...431..963S} code are very demanding in computing time.
\citet{MGV00} use a static 3D photoionization code. The  diffuse radiation treated in the on-the-spot (OTS approximation) -- similarly to the former two groups. They apply their tool  to a simple  case showing how velocity field and geometry go together. The main draw-back of this code is the absence of a full treatment of the diffuse field, despite the huge time of execution needed to compute a model.
\citet{HBRH03} used a morphological modelling code to help determine the geometry (set to axisymmetric), structure and kinematics (a linear expansion law is used) of the ellipsoidal PN Sa 2-21. The code SHAPE  \citep{2006astro.ph..1585S}, which works in 3D, is very versatile concerning the exploration of line profiles, but, as for the code of \citet{HBRH03}, the distribution of the emissivities is arbitrary and do not come from a self-consistent, photoionization approach. It must use the output of a 3D photoioinization code to improve the emissivities distribution. The main drawback of the actual version of SHAPE is the use of a commercial 3-D modeling software developped to run under the Microsoft operating system.
\citet[][and references therein]{2006astro.ph..1283S} try to solve the long-standing problem of ``deprojecting the bi-dimensional apparent morphology of a three-dimensional mass of gas''. They use the zero-velocity-pixel-column information to constrain 1D photoionization models and the central-star-pixel-line to constrain the expansion velocity of the emitting gas. Those technics lead to a possible reconstruction of observed nebulae in 3D. \citet[][and references therein]{STCB04} use tomography to reconstruct the morphologies of PNs, but are limited by the assumptions of linear expansion laws.
\citet[][and references therein for the same group]{2006astro.ph..1439G} use a 1D static photoionization code under the OTS assumption to  compute emission line profiles and compare them to observations of about 100 planetary nebulae. They allow for complex velocity fields, but are severely limited by the spherical geometry. 
The only fully 3D photoionization codes with accurate transfer are MOCASSIN \citep{Erc03} and the code of \citet{WME04}. There is no computation of line profiles yet, but VISNEB can be applied to them            

Ideally, of course, one would like fully hydrodynamical, 3D models including
photoionization. Even when this exists, static pseudo-3D models such as ours
will remain useful, because of their relative simplicity, to explore the
parameter space. Of the codes explicitly designed for comparison with data, the cunjunction of NEBU\_3D with VISNEB so far appears as the best compromise between relevance, versatility and rapidity.}
             \section{Complex profiles with a simple expansion law}
\label{sec:appli1}

             \subsection{A bipolar nebula toy model}
\label{sub:bipol}

\begin{figure}
\begin{center}
\includegraphics[width=9.cm]{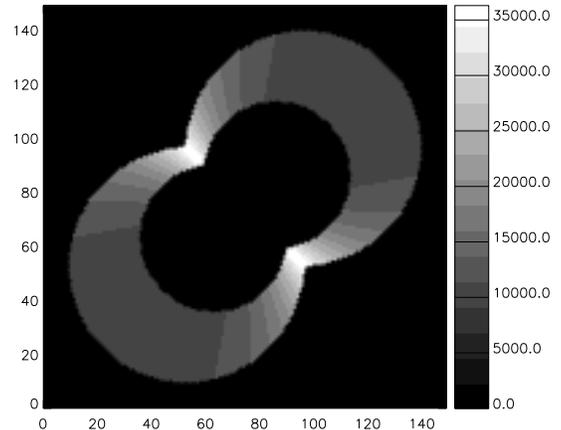}
\end{center}
\caption{Electron density distribution, for the 
bipolar model described in Sect 3.1, in which the 
hydrogen  density is kept constant along any 
radial direction.\label{fig:shape-de1}}
\end{figure}

The photoionization model presented in this 
section consists in applying \nebu/ to a set of 
40 runs of NEBU, each one corresponding to a 
different polar angular distance $\theta$ and 
characterized by a different gas density 
distribution. We consider no azimuthal variation 
for the density distribution. Since the model is 
axisymmetric, we need to model only 1/8th of the 
nebula, obtaining the rest of it by rotation. 
Thus, we vary $\theta$ from 0 to $\pi/2$, from 
the equator to the pole.

The inner radius,  $R_\mathrm{inner}(\theta)$, is 
defined in such a way that the inner cavity has a 
bipolar shape, delineated by two equal 
intersecting spheres of radius R$_{2}$. The 
center of each sphere is displaced on the polar 
axes by a distance R$_{1}$ from the ionizing 
central source. The equation for the inner radius 
is then:
$$R_\mathrm{inner}(\theta)= \sqrt{R_2^{2} 
-R_1^{2} \cos^{2}(\theta)}+R_1 \sin(\theta)$$

We assume that the hydrogen density at the inner 
surface is related to the distance of this 
surface to the ionizing star by:
$n_\mathrm{inner}(\theta) = 
n_\mathrm{equat}\times(R_\mathrm{equat}/R_\mathrm{inner}(\theta))^{2}$. 
For each polar angle $\theta$, the variation of 
the hydrogen density along the radius is given by:
$n(R,\theta)_{R > R_\mathrm{inner}(\theta)} = 
n_\mathrm{inner}(\theta)\times(R_\mathrm{inner}(\theta)/R)^\gamma$ 
(n. b.  $\gamma = 0$ corresponds to a density 
that is constant along each direction, $\gamma=2$ 
corresponds to a free expansion in each 
direction).
The free parameters for these geometries are then 
: $R_{1}/R_{2}, 
R_\mathrm{equat},n_\mathrm{equat}$ and $\gamma$. 
For the model presented in this section, we fix 
$R_{1}/R_{2}$ so that the equatorial inner size 
of the nebula is half the polar inner size. This 
assumption implies $3 R_{2} = 5 R_{1}$. We also 
take $\gamma$=0, and fix $R_\mathrm{equat}$ and 
$n_\mathrm{equat}$ so that the size of the shell 
is of the order of the size of the inner cavity, 
namely $R_\mathrm{equat}= 4 \times 10^{16}$~cm 
and $n_\mathrm{equat} = 3 \times 
10^{4}$~cm$^{-3}$.

The stellar energy distribution is that of a 
blackbody with \teff/=70~kK. We adopt a stellar 
luminosity of  L$_*$=1$\times 10^{37}$~erg 
s$^{-1}$. The elemental abundances are taken 
solar. Dust is not included in the model.

The interpolation is done in a cube of 100$^3$ 
pixels, each of 5$\times 10^{15}$ cm in size. At 
an assumed distance of 2.7~kpc from the observer, 
this corresponds to an angle of 0.075\arcsec for 
each pixel.

We show in Fig.~\ref{fig:shape-de1} a cut in a 
plane containing the polar axis, with the 
electron density represented by different levels 
of grey. Black zones correspond to null electron 
density (either because of lack of matter like in 
the inner region, or due to the fact that no 
ionizing photons reach this zone, like in the 
outer parts). From the simple assumptions used to 
define the morphology of the nebula, it follows 
that the shape of the recombination front is 
bipolar, with a strong enhancement of the density 
at the equatorial ring.

             \subsection{Computed monochromatic images and line profiles}

\begin{figure*}
\begin{center}
\includegraphics{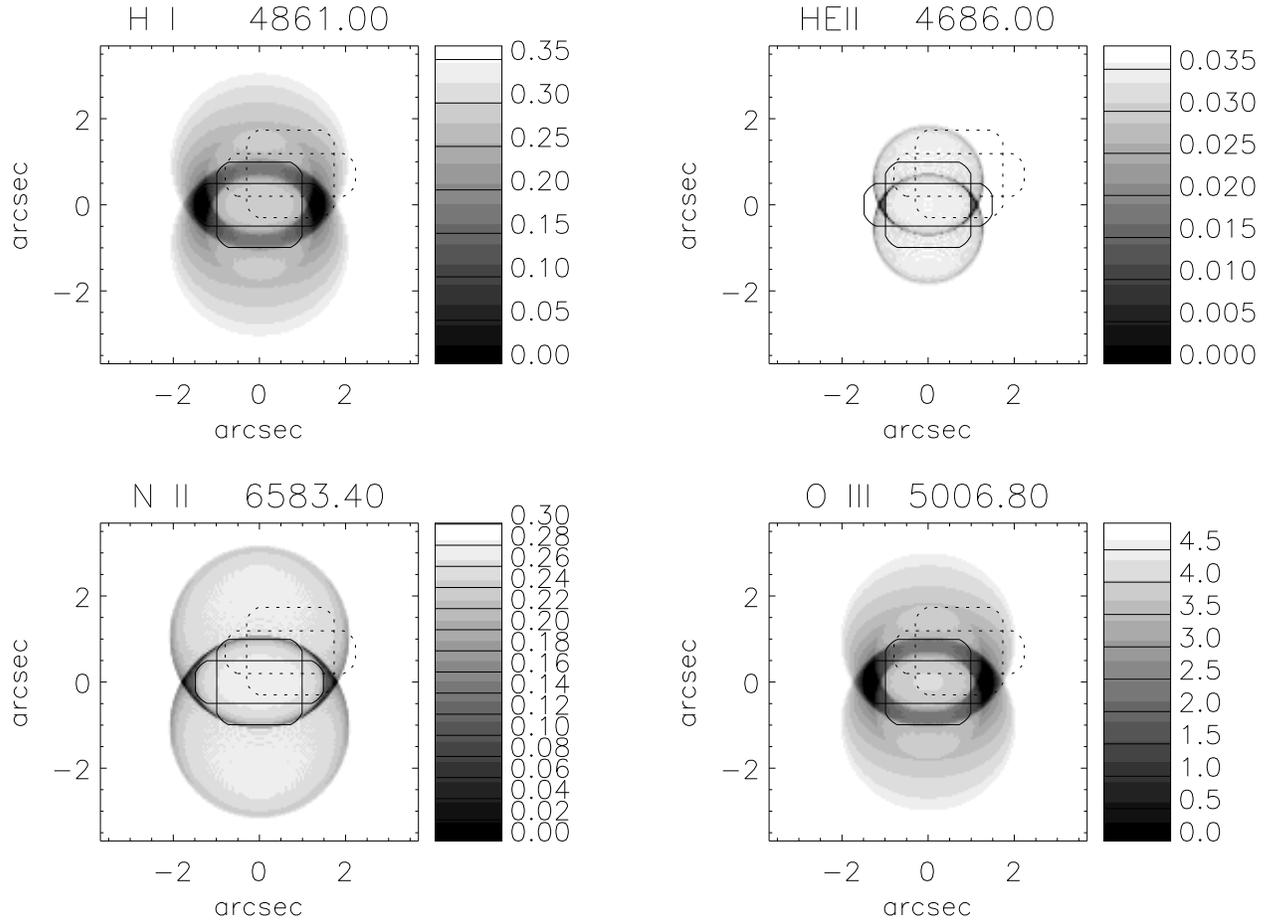}
\end{center}
\caption{Monochromatic images obtained for three 
emission lines (\hbeta/, \alloa{He}{ii}{4689}, 
\forba{N}{ii}{6583}, and \forba{O}{iii}{5007}), 
for the bipolar nebula shown in 
Fig.~\ref{fig:shape-de1} and described in Sec. 
3.1. The apertures used to compute the emission 
line profiles shown in Fig.~\ref{fig:resPro1} are 
superimposed to the images: the sizes of the 
apertures are: 1\arcsec x3\arcsec\- and 2\arcsec 
x2\arcsec. The largest aperture (10\arcsec 
x10\arcsec) covers the entire nebula and is not 
shown. The centered apertures are drawn with 
solid lines. Off-center apertures  are drawn with 
dashed lines. Surface brightness units are 
arbitrary, but the same for the four 
images.\label{fig:resIm1}}
\end{figure*}

\begin{figure*}
\begin{center}
\includegraphics{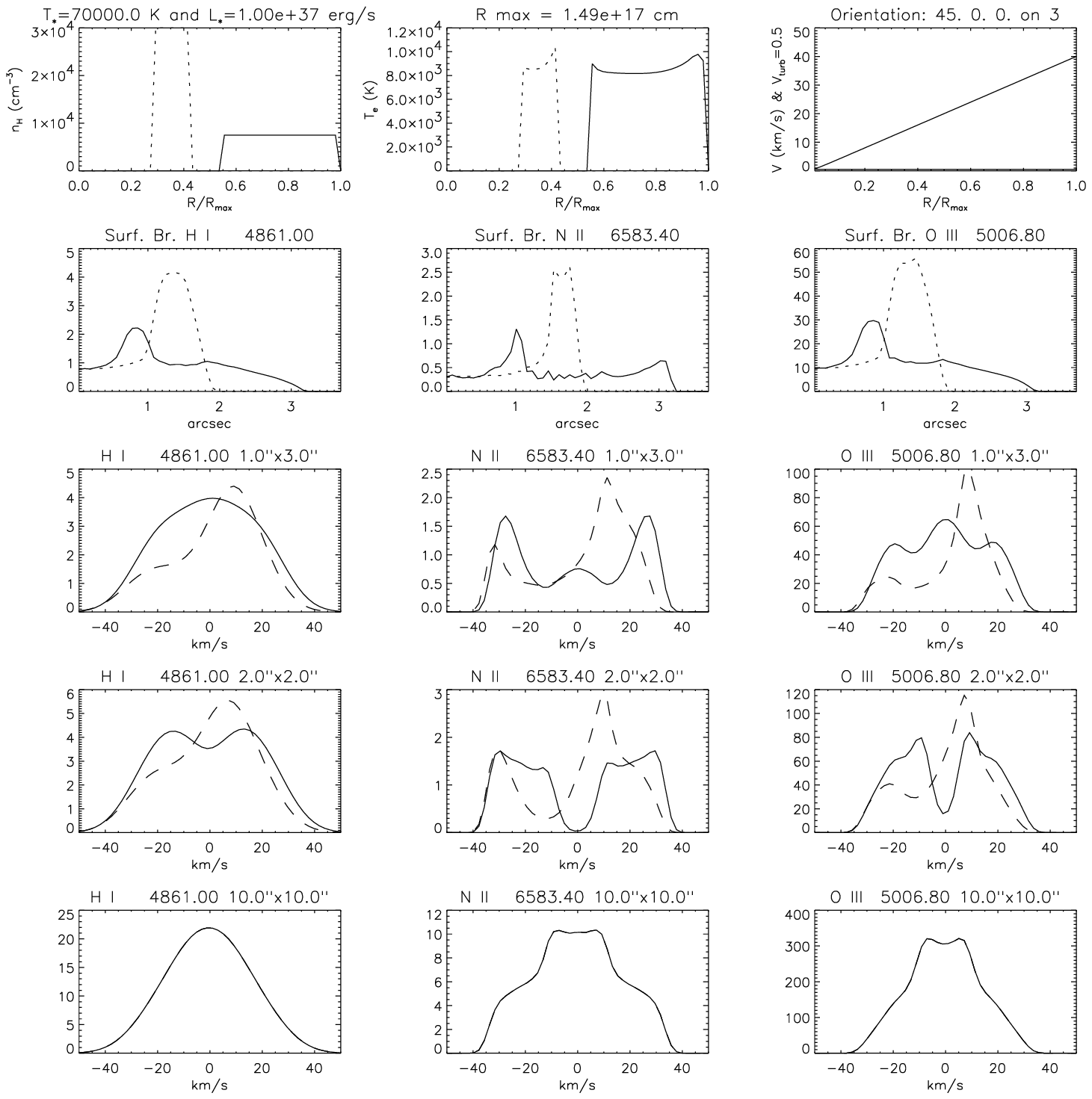}
\end{center}
\caption{Some physical parameters, surface 
brightness distribution and line profiles for the 
bipolar nebula shown in Fig.~\ref{fig:shape-de1} 
and described in Sec. 3.1. Upper row:
radial distribution of the hydrogen density (left 
panel), the electron temperature (middle panel) 
and the expansion velocity (right panel). Solid 
curves correspond to the polar direction, dotted 
curves to a perpendicular one. Second row : from 
left to right, surface brightness distribution in 
the  \hbeta/, \forba{N}{ii}{6583} and 
\forba{O}{iii}{5007} lines along the polar axis 
(solid curves) and in the perpendicular direction 
(dotted curves).
Last three rows: from left to right, line 
profiles \hbeta/, \forba{N}{ii}{6583} and 
\forba{O}{iii}{5007} lines, through apertures 
represented in Fig.~\ref{fig:resIm1} (the size of 
the aperture is specified on top of each plot). 
Solid curves correspond to centered apertures, 
dashed curves to off-centered apertures. 
Intensity units are arbitrary, but the same for 
all the plots.
\label{fig:resPro1}}
\end{figure*}

The images presented in Fig.~\ref{fig:resIm1} are 
computed for a nebula with a  polar axis making 
an angle of 45degr with the plane of the sky. A 
seeing of 0.5\arcsec\- is assumed.
The figure shows the surface brightness of the 
nebula in four emission lines:  \hbeta/, 
\forba{O}{iii}{5007}, \forba{N}{ii}{6583} and 
\alloa{He}{ii}{4686}. The bipolar shape is seen 
in all the lines, but the images differ due to 
the ionization structure of the nebula. 
Obviously, it is the \forba{N}{ii}{6583} image 
which best suggests the presence of the inner 
cavity. \news{There is a small ``noise'' (the classical 
``Moir\'e'' effect) due to the finite size of the 
cells. This effect is also present in the surface 
brightness profiles shown in the next sections. 
This effect is more important for emission lines 
that are occupying a thin shell than for HI line 
for example. It decreases when the number of 
cells of the main cube increases.}

The emission line profiles are computed for 
various apertures, indicated in 
Fig.~\ref{fig:resIm1}. Solid lines correspond to 
centered apertures, dashed lines correspond to 
off-center apertures. The off-center apertures 
were shifted  by 0.3\arcsec\- and 0.5\arcsec\- in 
the x- and y-directions respectively. The 
proportion of the \hbeta/ flux collected through 
the 1\arcsec x3\arcsec\- and the 2\arcsec 
x2\arcsec\- centered apertures are 18 and 19\% 
respectively.

To compute the line profiles, we adopted a very 
simple expansion velocity law, given by the 
following expression:  $\vec{V}(\vec{R}) = 
V_0.\vec{R}/R_{max}$, where  $\vec{R}$ is the 
position vector originating at the centre of the 
nebula, $R_{max}$ is maximum distance of the 
ionization front to the centre (i.e. the radius 
in the polar direction), and  $V_0$ is set to 40 
\kms/.

The emission line profiles are computed on a 60 
pixels grid (from -50 to 50 \kms/).

The results are shown in Fig.~\ref{fig:resPro1} 
for three lines:  \hbeta/, \forba{N}{ii}{6583} 
and \forba{O}{iii}{5007} (respectively left, 
middle and right panels of the three bottom 
rows). The shape and size of the slit is 
indicated at the top of each panel. Profiles 
obtained through off-center slits are represented 
by dashed lines. The lowest row of panels is for 
an aperture that covers the entire nebula. This 
case  corresponds to what would be observed for 
an extragalactic planetary nebula, for example. 
The figure also shows some plots that are useful 
to understand the line profiles. The top  panels 
give the  radial distribution of the hydrogen 
density, the electron temperature and the 
expansion velocity. Solid curves correspond to 
the polar direction, dotted curves correspond to 
a direction perpendicular to it. The second row 
of panels shows the surface brightness 
distribution in the  \hbeta/, \forba{N}{ii}{6583} 
and \forba{O}{iii}{5007} lines along the polar 
axis (solid curves) and in the perpendicular 
direction (dotted curves).

As expected, the lines showing most structure are 
the  \forba{N}{ii}{6583} ones, which are emitted 
in the outskirts of the nebula, and sampling the 
largest variety of radial velocities. Note that 
the centered square 2\arcsec x2\arcsec\ aperture 
misses the zone of zero radial velocity, i.e. the 
zones where the velocity vector is entirely in 
the plane of the sky, while the rectangular 
aperture does not, leading to a very different 
line profile. In the case of the 
\forba{O}{iii}{5007} line, the rectangular slit 
actually intercepts a large zone with zero radial 
velocity, resulting in a profile without 
line-splitting. The \hbeta/ line profile is the 
most featureless, essentially because of the 
strong effect of thermal velocity on this line. 
The important difference between the profiles of 
lines emitted by various ions amply justifies the 
use of a photoionization code to analyze the 
kinematics of real nebulae.

The position of the slit with respect to the 
center of the nebula has a dramatic influence on 
the observed line profiles, as can be seen by 
comparing the dashed and solid curves in 
Fig.~\ref{fig:resPro1}. This means that a 
perfectly symmetric nebula can be mistaken for a 
nebula with uneven density distribution or 
complex velocity field simply because the slit is 
not perfectly centered!

             \section{Mimicking the effect of a turbulent velocity field}
\label{sec:turbulence}

\begin{figure*}
\begin{center}
\includegraphics{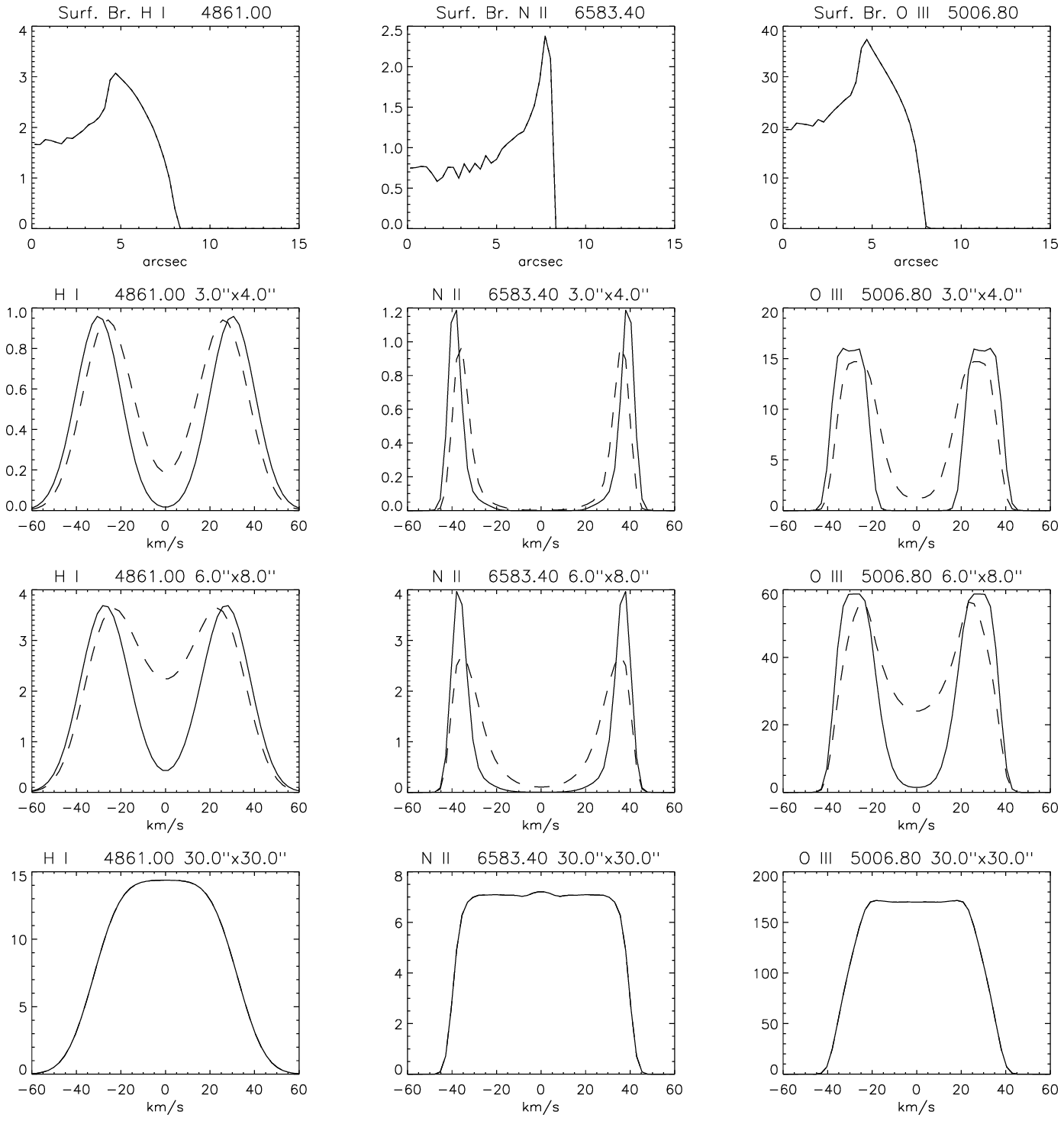}
\end{center}
\caption{Spherical nebula, with a Hubble flow 
expansion law and with negligible turbulence. 
The setup of the figure is the same as for rows 2 
-- 5  of  Fig.~\ref{fig:resPro1}.
   }
\label{fig:resSph1}
\end{figure*}

\begin{figure*}
\begin{center}
\includegraphics{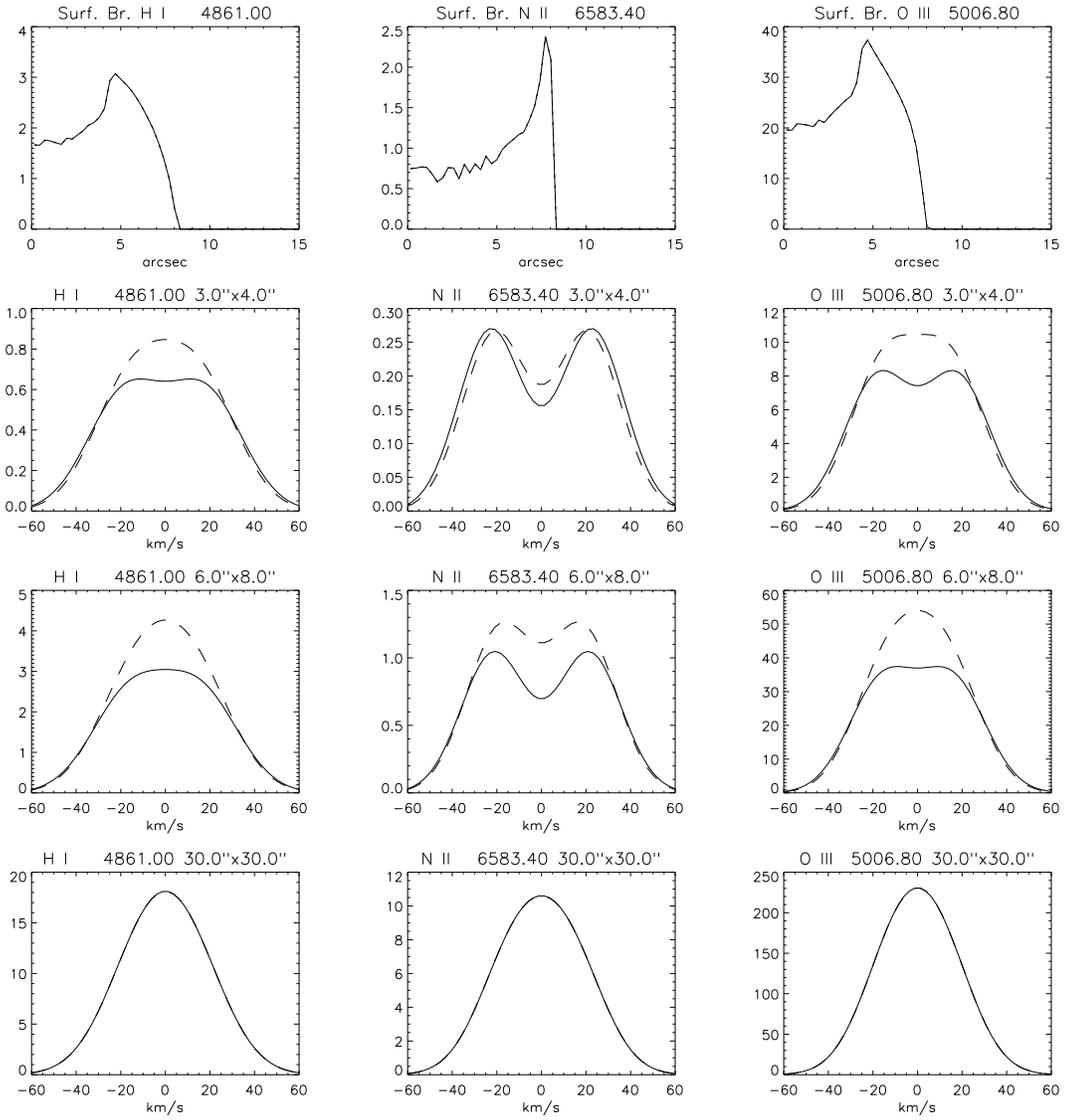}
\end{center}
\caption{Spherical nebula, with a low velocity 
Hubble flow expansion law and high turbulence. 
The setup of the figure is the same as for 
Fig.~\ref{fig:resSph1}. }
\label{fig:resSph2}
\end{figure*}

\begin{figure*}
\begin{center}
\includegraphics{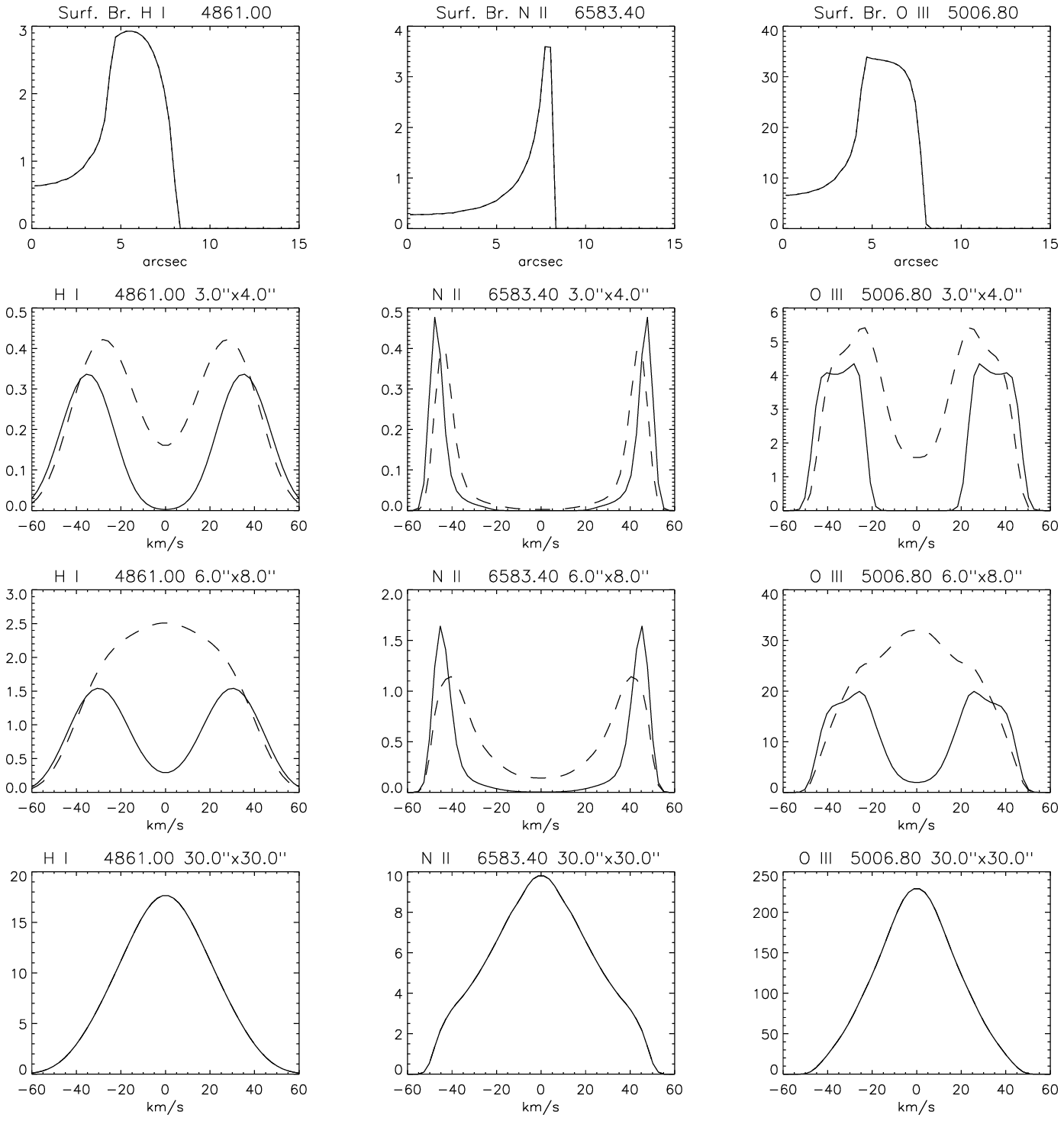}
\end{center}
\caption{Ellipsoidal nebula, seen pole-on, with a 
Hubble flow expansion law and without turbulence. 
The setup of the figure is the same as for 
Fig.~\ref{fig:resSph1}.
\label{fig:resEli1}}
\end{figure*}

Recently, \citet{NAGS00} and \citet{GAZ03} have 
claimed to find evidence for turbulence in 
certain planetary nebulae, especially in those 
ionized by [WC] type central stars. While it is 
quite possible that the strong wind from [WC] 
stars induces turbulent motions in the nebular 
gas \citep[see][]{2003IAUS..209..507M}, we feel 
that the observational evidence is not as strong 
as might be thought.

The effect of turbulence is to broaden the 
profiles by  the same amount for  lines of all 
elements. In extreme cases, this may result in 
the line splitting to disappear, if the 
broadening by turbulence is important. This is 
illustrated in Figs.~\ref{fig:resSph1} and 
~\ref{fig:resSph2}, which show the results from 
two spherical models. Both models are defined by 
an ionizing star of \teff/=70~kK and 
L$_*$=1$\times 10^{37}$~erg s$^{-1}$ surrounded 
by a spherical nebula with  an inner cavity of 
7.5$\times 10^{16}$cm and a constant hydrogen 
density of 8000 cm$^{-3}$. The first model has a 
linear expansion law reaching a value of 40~\kms/ 
at the recombination front, while the second one 
has a turbulence of 20~\kms/ added and the 
velocity at the recombination front is reduced to 
25~\kms/. In order to construct the surface 
brightness and velocity profile figures, a 
distance to the object of 1.16~kpc has been 
assumed, implying an outer angular radius of 
8~\arcsec. The setup of the figures is the same 
as for Fig.~\ref{fig:resPro1} except that the 
first row has been omitted. We see in 
Fig.~\ref{fig:resSph2} that turbulence has 
strongly reduced the two peaks that were seen 
with the 3\arcsec x4\arcsec\ slit.

There are however several other ways to achieve 
such a broadening. Even in spherical symmetry, 
one can produce  similar profiles by changing at 
the same time the density distribution and the 
expansion velocity field. Another way is to 
consider an ellipsoidal nebula, rather than a 
spherical one, and oriented pole-on. With such a 
configuration, the weight of the zones of nearly 
zero radial velocity becomes large, and high 
velocity wings are produced as well. 
Fig.~\ref{fig:resEli1} shows such a case, which 
differs from the case shown in 
Fig.~\ref{fig:resSph1} in that the nebula is now 
ellipsoidal with a cavity of 7.5$\times 
10^{16}$cm and $1.12\times 10^{17}$cm and an 
inner density of 8000 cm$^{-3}$ and 3550 
cm$^{-3}$, in the equatorial and polar directions 
respectively. The expansion law is identical and 
reaches a maximum velocity at the polar 
ionization front of 70~\kms/. As seen in 
Fig.~\ref{fig:resEli1}, this ellipsoidal model 
has very similar line profiles to the turbulent 
spherical model, at least if the slit size covers 
the whole nebula (and for \hbeta/, even if the 
slit covers a substatial part of it and is 
off-center). Note that even the surface 
brightness profiles are very similar. Therefore, 
nebulae that would correspond to such models 
would be impossible to distinguish in practise, 
if using observational setups such as those 
represented by the last two rows of 
Figs.~\ref{fig:resSph1} and ~\ref{fig:resEli1}. 
Note, however, that with a smaller slit, the two 
models can be distinguished, at least if the slit 
is perfectly centered (compare the second rows of 
Figs.~\ref{fig:resSph1} and ~\ref{fig:resEli1}) 
because in that case the splitting of the 
emission lines due to crossing two emitting 
regions on the line of sight.

             \section{Distinguishing a blister 
from its spherical impostor using line profiles}
\label{sec:blister}

\begin{figure*}
\begin{center}
\leavevmode
\includegraphics{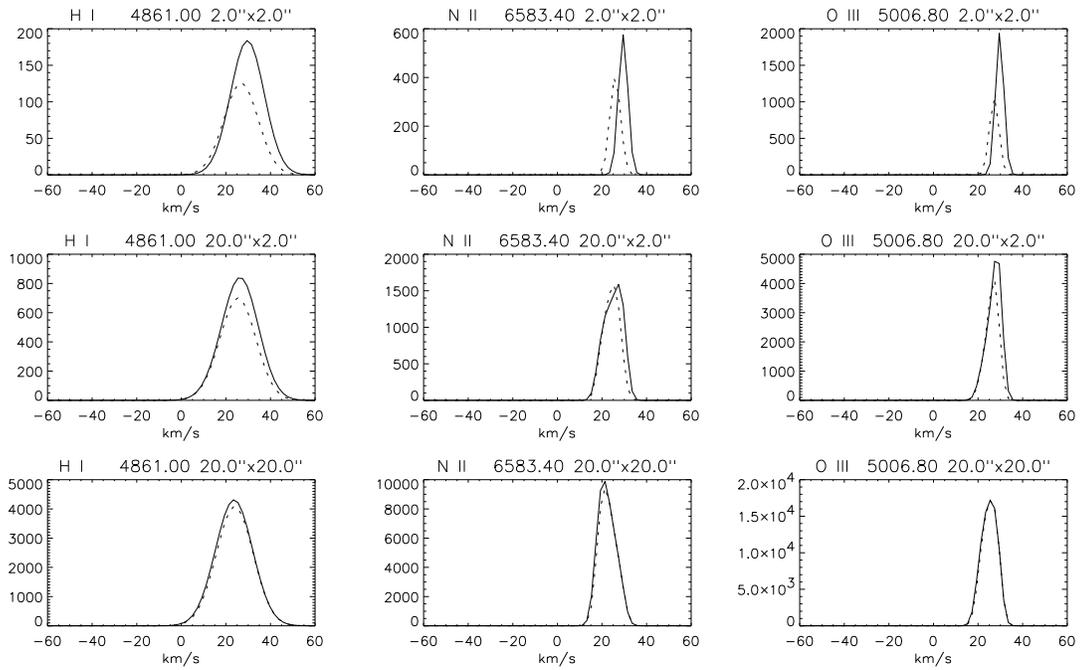}
\end{center}
\caption{Line profiles for the face-on blister 
(see Sect.~\ref{sec:blister}.  The setup of the 
figure is the same asfor rows 3 -- 5   of 
Fig.~\ref{fig:resPro1}. }\label{fig:resBli}
\end{figure*}

\begin{figure*}
\begin{center}
\leavevmode
\includegraphics{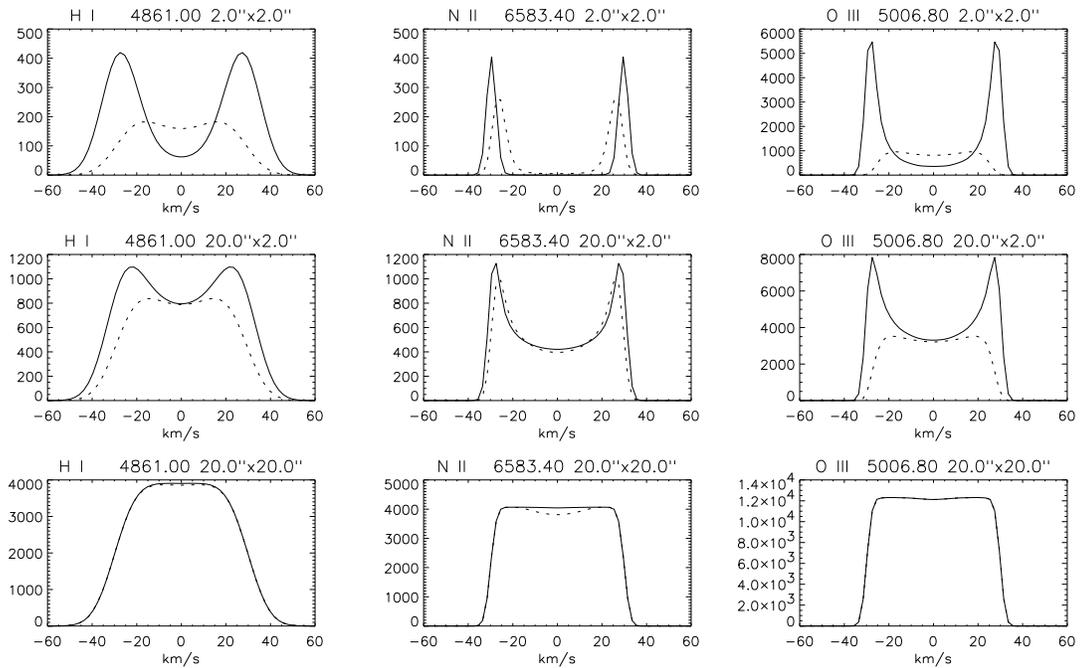}
\end{center}
\caption{Line profiles for the spherical impostor 
(see Sect.~\ref{sec:blister}).  The setup of the 
figure is the same as for rows 3 -- 5 of 
Fig.~\ref{fig:resPro1}.} \label{fig:resSph}
\end{figure*}

In Paper~I we showed an example where  two 
totally different geometries could lead to the 
same surface brightness \hbeta/ maps: a face-on 
blister could be mistaken for a sphere -- which 
we named the ``spherical impostor''\footnote {In 
the right panel of Fig. 17 in Paper~I, the line 
styles to represent the blister and the spherical 
models have been inverted: it is the blister 
model which was represented by a continuous line, 
contrary to what is stated in the legend and to 
what has been done for the other panels of the 
figure}.  Here we illustrate how high resolution 
spectroscopy can allow one to discriminate 
between the two geometries. Indeed, the velocity 
fields for these two geometries are expected to 
be very different. For the face-on blister, the 
gas is streaming out of the neutral cloud towards 
the observer, while for the spherical impostor, 
one expects a spherically symmetric expansion.

Fig.~\ref{fig:resBli} shows the line profiles of 
the blister model of Paper~I, in which we have 
assumed that the velocity is perpendicular to the 
gas surface and is set to 30~\kms/. 
Fig.~\ref{fig:resSph} shows the line profiles of 
the spherical impostor model from Paper~I, in 
which we have assumed a constant velocity of 
30~\kms/.
The differences between the observed profiles 
between Fig.~\ref{fig:resBli} and 
Fig.~\ref{fig:resSph} are spectacular. For the 
blister, all the lines are blue-shifted and 
narrow, since gas flows in only one direction.
For the spherical impostor, when using a thin 
slit passing through the center, the emission 
lines are splitted. If the slit covers a large 
fraction of the object or if it is off center, 
there is no splitting, but the lines are broad 
(their width at the base is twice the expansion 
velocity).

             \section{Generating 3D maps and PV-diagramms}

\begin{figure*}
\begin{center}
\includegraphics{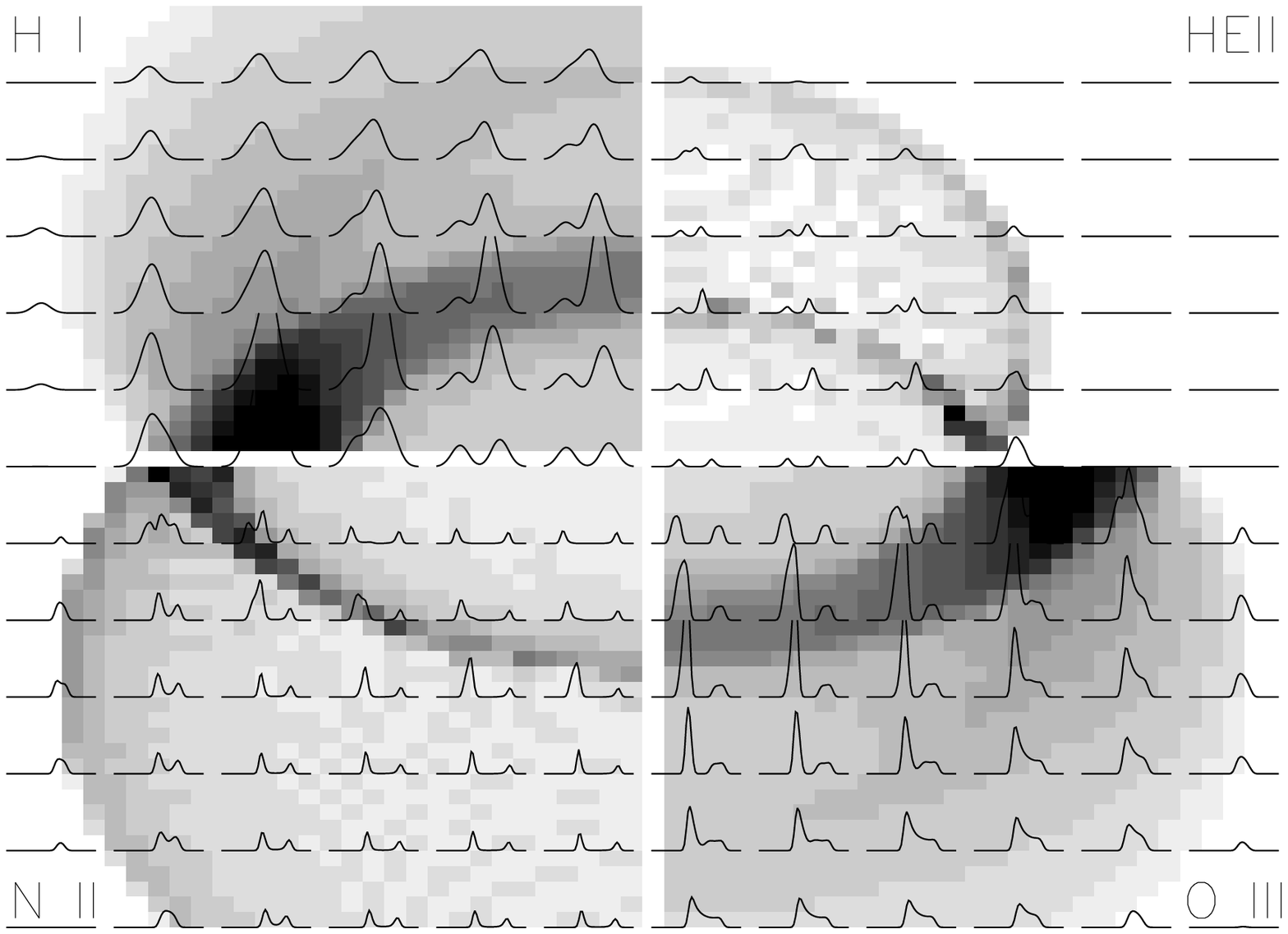}
\end{center}
\caption{A 3D map of the bipolar nebula model 
from Sec.~\ref{sub:bipol}. Each quadrant 
corresponds to a different emission line. The 
surface brightness is represented by levels of 
grey. Superimposed on this image are the profiles 
of the same lines, integrated over areas of 5x5 
pixels. \label{fig:IM1}}
\end{figure*}

From examples shown in the previous sections, it 
is clear that the interpretation of line profiles 
is by no means straightforward. One of the 
problems is that observed profiles strongly 
depend on the position and size of the slit. 
Obviously, a large amount of information is 
necessary to unambiguously unveil the morphology 
and kinematics of a nebula. There are now an 
increasing number of instruments which use 
integral field units (IFU) to achieve 
spectroscopy of an extended portion of the sky, 
either using lens arrays, optical fibres or image 
slicers. These instruments can thus generate 3D 
maps of extended objects, with wavelength (or 
velocity) being the third dimension. Such 
examples can be found in 
\citet{2004AJ....127.2145A,2005AJ....130.1707V}. 
Our tool is well-suited to also produce such maps 
for photoionization models of asymmetric nebulae. 
One such example is shown in Fig.~\ref{fig:IM1}, 
which represents  the bipolar nebula model from 
Sec.~\ref{sub:bipol}. Each quadrant corresponds 
to a different emission line (\hbeta/, 
\alloa{He}{ii}{4689}, \forba{N}{ii}{6583}, and 
\forba{O}{iii}{5007}). The grey-colour image 
represents the surface brightness in the line, 
computed with pixels of 0.075\arcsec x 
0.075\arcsec. Superimposed on this image are the 
profiles of the same lines, integrated over areas 
of 5x5 pixels. The line intensities are all on 
the same scale. Apart from illustrating how the 
line profiles depend on the emitting ion, this 
figure also clearly shows how the profiles change 
with position. In the present case, they may 
change from single-peak to triple-peak 
structures. The latter occur near the waist of 
the bipolar structure, when, due to inclination, 
the line of sight crosses the front and back of 
one bubble, plus the front on the other bubble 
(see Fig.~\ref{fig:shape-de1} for visualization). 
Note that the largest component of the 
double-peak line profiles is the red one in the 
upper spectra and the blue one in the lower 
spectra because of the inclination of the nebula.

\begin{figure*}
\begin{center}
\includegraphics{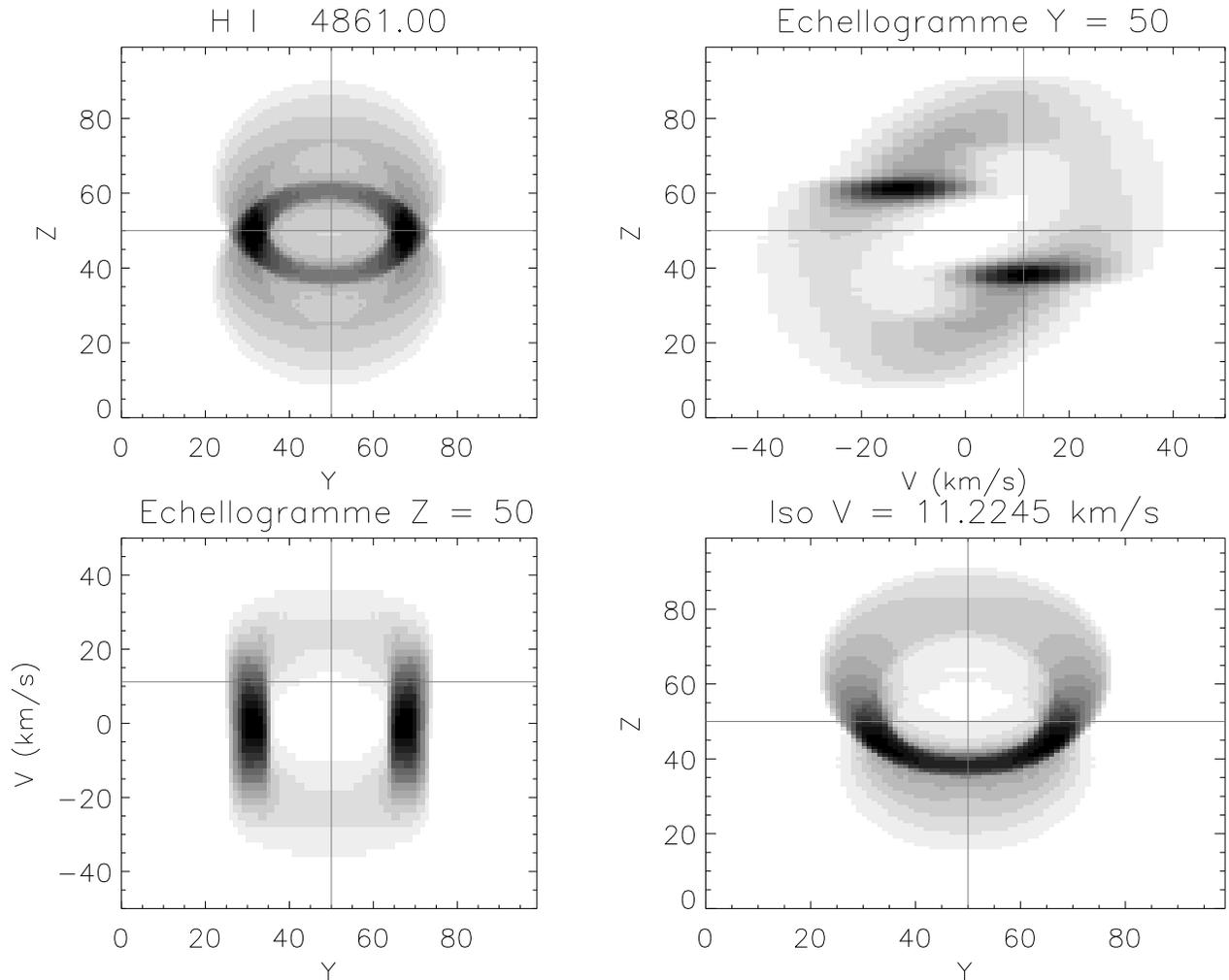}
\end{center}
\caption{PV diagrams obtained for the bipolar 
nebula presented in Sec.~\ref{sub:bipol}.Upper 
left: surface brightness image for \hbeta/ with 
the 2 slits used for the forthcoming PV diagrams, 
Upper right: PV diagram obtained for a narrow 
vertical slit crossing the center of the nebula, 
Lower left: PV diagram obtained for a narrow 
horizontal slit crossing the center of the 
nebula, Lower right: velocity channel map 
obtained from the V=11~\kms/ channel. 
\label{fig:PV}}
\end{figure*}

To illustrate the capacities of VELNEB\_3D to 
produce Position-Velocity (PV) diagrams, the 
bipolar model described in Sec.~\ref{sub:bipol} 
is used to generate images presented in 
Fig.~\ref{fig:PV}. The \hbeta/ surface brightness 
map is shown with the two narrow slits 
superimposed, which are  used to compute the 
PV-diagrams. One is obtained for an horizontal 
slit, the second one for a vertical one. They are 
both centered, but any position can be defined by 
the user. The nebula has its polar axis making an 
angle of 45degr with the plane of the sky, 
resulting in well defined red and blue components 
for the vertical slit (upper right panel of 
Fig.~\ref{fig:PV}), when the slit is crossing the 
two bubbles of the nebula. This orientation 
effect is not seen in the case of the horizontal 
slit crossing mainly gas at projected velocity 
closed to zero (lower left panel of 
Fig.~\ref{fig:PV}).
The last panel shows a velocity channel map 
obtained for velocity close to 11~\kms/. The part 
of the nebula expanding in the positive direction 
is well traced.
             \section{Summary and prospects}

In this paper, we have presented a tool, 
VELNEB\_3D, which uses the results from any 3D 
photoionization model to generate emission line 
profiles, position-velocity maps and 3D maps in 
any emission line by assuming an arbitrary 
velocity field. In this work, the code used to 
generate the photoionization model was NEBU\_3D 
\citep{MSP05a} but any other code can be used.

So far, most of of the modelling of line profiles 
in planetary nebulae has been done using 1D codes 
\citep[see][and references therein]{GAZ03}. The 
examples given in this paper show that by taking 
into account deviation from spherical symmetry, 
the interpretation on the velocity field may be 
different. This is important, since about 80\% of 
planetary nebulae are not round \citep{Balick06}.

With this tool, we have been able to show how 
much the interpretation of observed line profiles 
may depend on the exact position of the slit. For 
example, a complex line profile may be obtained 
even with a simple expansion law if  the nebula 
is not spherical and the slit slightly 
off-center. Since physically, geometry and 
velocity field are related (the geometry being 
the result of the evolution of the velocity field 
in time) it is likely that in real nebulae with 
complex profiles, both the geometry and the 
velocity field are in fact complex. Our tool 
allows one to explore a larger parameter space 
than 1D models and (perhaps) pin down the best 
solution for observed nebulae. It can also be 
useful as a tutorial kit for those wishing to 
better understand observed monochromatic images 
and line profiles.

We believe that trying to reproduce observations 
of real ionized nebulae with our tool should give 
more insight into the physics of these objects 
and guide the evolutionary dynamical modelling of 
planetary nebulae such as done by e.g. 
\citet{2002ApJ...581.1204V,2005AA...431..963S} in 
1D and 
\citet{1993ApJ...404L..25F,1995MNRAS.277..173M,1997AA...321L..29M,2003ApJ...585L..49V} 
in 2D.

While dynamical modelling is obviously the 
ultimate  step in understanding the origin and 
evolution of planetary nebula shapes, or of the 
shapes of other kinds of nebulae, it involves 
heavy computation and at the same time a 
simplification of the physics at work. Therefore 
an exploration with a tool such as we propose is 
worthwhile.

Our examples suggest that it may be actually very 
difficult to reconstruct the geometry and 
velocity field of any real nebula without any 
underlying assumptions. However, there are cases 
where high resolution spectroscopy may 
distinguish between two very different 
geometries. Such is the case of a face-on blister 
and its spherical impostor, which have the same 
surface brightness distribution in H$\alpha$ for 
the same ionizing star.

Our tool may also be useful in preparing and 
interpreting observations of ionized nebulae with 
integral field units, which are able to provide 
3D maps and will predictably lead to a major 
progress in nebular astronomy in the coming 
years. We foresee however the difficulty that 
with such a huge amount of observational and 
computational data, it may be tough to pin down 
the really important physics, in other words, to 
not only answer the question ``what is it that we 
see?'' but also ``why is it like that?'' Clearly, 
ways will have to be found to make the best use 
of these tools.

\news{Another application of our tool is the 
possibility to compute a large amount of models, 
varying the morphology, the ionization parameter, 
the velocity law and the position of the slit and 
to include the resulting emission line profiles 
in a Virtual Observatory (VO). The users of the 
VO would be able to quickly scan the catalog to 
search for all the nebulae satisfying given 
criteria, for example on line profiles.  Such a 
work is in progress.

The VELNEB\_3D tool is available on request to C.M.}

\acknowledgements

G.S. is grateful to the Instituto de Astronomia, 
UNAM, Mexico, for hospitality and financial 
support. The computations were carried out on a 
AMD-64bit computer financed by grant PAPIIT 
IX125304 from DGAPA (UNAM,Mexico). C.M. is partly 
supported by grant Conacyt-40095 (Mexico).


\end{document}